\documentstyle[twoside]{article}

\catcode`\@=11
\long\def\@makefntext#1{
\protect\noindent \hbox to 3.2pt {\hskip-.9pt
$^{{\eightrm\@thefnmark}}$\hfil}#1\hfill}               

\def\@makefnmark{\hbox to 0pt{$^{\@thefnmark}$\hss}}    

\def\ps@myheadings{\let\@mkboth\@gobbletwo
\def\@oddhead{\hbox{}
\rightmark\hfil\eightrm\thepage}
\def\@oddfoot{}\def\@evenhead{\eightrm\thepage\hfil
\leftmark\hbox{}}\def\@evenfoot{}
\def\sectionmark##1{}\def\subsectionmark##1{}}

\oddsidemargin=\evensidemargin
\newcounter{sectionc}\newcounter{subsectionc}\newcounter{subsubsectionc}
\renewcommand{\section}[1] {\vspace{12pt}\addtocounter{sectionc}{1}
\setcounter{subsectionc}{0}\setcounter{subsubsectionc}{0}\noindent
        {\tenbf\thesectionc. #1}\par\vspace{5pt}}
\renewcommand{\subsection}[1] {\vspace{12pt}\addtocounter{subsectionc}{1}
        \setcounter{subsubsectionc}{0}\noindent
        {\bf\thesectionc.\thesubsectionc. {\kern1pt \bfit #1}}\par\vspace{5pt}}
\renewcommand{\subsubsection}[1] {\vspace{12pt}\addtocounter{subsubsectionc}{1}
        \noindent{\tenrm\thesectionc.\thesubsectionc.\thesubsubsectionc.
        {\kern1pt \tenit #1}}\par\vspace{5pt}}
\newcommand{\nonumsection}[1] {\vspace{12pt}\noindent{\tenbf #1}
        \par\vspace{5pt}}

\newcounter{appendixc}
\newcounter{subappendixc}[appendixc]
\newcounter{subsubappendixc}[subappendixc]
\renewcommand{\thesubappendixc}{\Alph{appendixc}.\arabic{subappendixc}}
\renewcommand{\thesubsubappendixc}
        {\Alph{appendixc}.\arabic{subappendixc}.\arabic{subsubappendixc}}

\renewcommand{\appendix}[1] {\vspace{12pt}
        \refstepcounter{appendixc}
        \setcounter{figure}{0}
        \setcounter{table}{0}
        \setcounter{lemma}{0}
        \setcounter{theorem}{0}
        \setcounter{corollary}{0}
        \setcounter{definition}{0}
        \setcounter{equation}{0}
        \renewcommand{\thefigure}{\Alph{appendixc}.\arabic{figure}}
        \renewcommand{\thetable}{\Alph{appendixc}.\arabic{table}}
        \renewcommand{\theappendixc}{\Alph{appendixc}}
        \renewcommand{\thelemma}{\Alph{appendixc}.\arabic{lemma}}
        \renewcommand{\thetheorem}{\Alph{appendixc}.\arabic{theorem}}
        \renewcommand{\thedefinition}{\Alph{appendixc}.\arabic{definition}}
        \renewcommand{\thecorollary}{\Alph{appendixc}.\arabic{corollary}}
        \renewcommand{\theequation}{\Alph{appendixc}.\arabic{equation}}
        \noindent{\tenbf Appendix \theappendixc #1}\par\vspace{5pt}}
\newcommand{\subappendix}[1] {\vspace{12pt}
        \refstepcounter{subappendixc}
        \noindent{\bf Appendix \thesubappendixc. {\kern1pt \bfit #1}}
        \par\vspace{5pt}}
\newcommand{\subsubappendix}[1] {\vspace{12pt}
        \refstepcounter{subsubappendixc}
        \noindent{\rm Appendix \thesubsubappendixc. {\kern1pt \tenit #1}}
        \par\vspace{5pt}}

\newcommand{\textlineskip}{\baselineskip=13pt}
\newcommand{\smalllineskip}{\baselineskip=10pt}

\def\eightcirc{
\begin{picture}(0,0)
\put(4.4,1.8){\circle{6.5}}
\end{picture}}
\def\eightcopyright{\eightcirc\kern2.7pt\hbox{\eightrm c}}

\newcommand{\copyrightheading}[1]
        {\vspace*{-2.5cm}\smalllineskip{\flushleft
{\bf OUT--4102--57} \hfill         
{\bf MZ--TH/95--12}
\hfill{\bf UTAS--PHYS--95--12}     
\\}}

\def\abstracts#1#2#3{{
        \centering{\begin{minipage}{4.5in}\baselineskip=10pt\footnotesize
        \parindent=0pt #1\par
        \parindent=15pt #2\par
        \parindent=15pt #3
        \end{minipage}}\par}}

\newcommand{\bibit}{\nineit}
\newcommand{\bibbf}{\ninebf}
\renewenvironment{thebibliography}[1]
        {\frenchspacing
         \ninerm\baselineskip=11pt
         \begin{list}{\arabic{enumi}.}
        {\usecounter{enumi}\setlength{\parsep}{0pt}
         \setlength{\leftmargin 17pt}{\rightmargin 0pt}   
         \setlength{\itemsep}{0pt} \settowidth
        {\labelwidth}{#1.}\sloppy}}{\end{list}}

\newcounter{itemlistc}
\newcounter{romanlistc}
\newcounter{alphlistc}
\newcounter{arabiclistc}

\newcommand{\fcaption}[1]{
        \refstepcounter{figure}
        \setbox\@tempboxa = \hbox{\footnotesize Fig.~\thefigure. #1}
        \ifdim \wd\@tempboxa > 5in
           {\begin{center}
        \parbox{5in}{\footnotesize\smalllineskip Fig.~\thefigure. #1}
            \end{center}}
        \else
             {\begin{center}
             {\footnotesize Fig.~\thefigure. #1}
              \end{center}}
        \fi}

\newcommand{\tcaption}[1]{
        \refstepcounter{table}
        \setbox\@tempboxa = \hbox{\footnotesize Table~\thetable. #1}
        \ifdim \wd\@tempboxa > 5in
           {\begin{center}
        \parbox{5in}{\footnotesize\smalllineskip Table~\thetable. #1}
            \end{center}}
        \else
             {\begin{center}
             {\footnotesize Table~\thetable. #1}
              \end{center}}
        \fi}

\def\@citex[#1]#2{\if@filesw\immediate\write\@auxout
        {\string\citation{#2}}\fi
\def\@citea{}\@cite{\@for\@citeb:=#2\do
        {\@citea\def\@citea{,}\@ifundefined
        {b@\@citeb}{{\bf ?}\@warning
        {Citation `\@citeb' on page \thepage \space undefined}}
        {\csname b@\@citeb\endcsname}}}{#1}}

\newif\if@cghi
\def\cite{\@cghitrue\@ifnextchar [{\@tempswatrue
        \@citex}{\@tempswafalse\@citex[]}}
\def\citelow{\@cghifalse\@ifnextchar [{\@tempswatrue
        \@citex}{\@tempswafalse\@citex[]}}
\def\@cite#1#2{{$\null^{#1}$\if@tempswa\typeout
        {IJCGA warning: optional citation argument
        ignored: `#2'} \fi}}

\def\pmb#1{\setbox0=\hbox{#1}
        \kern-.025em\copy0\kern-\wd0
        \kern.05em\copy0\kern-\wd0
        \kern-.025em\raise.0433em\box0}

\def\fnt#1#2{\footnotetext{\kern-.3em
        {$^{\mbox{\scriptsize #1}}$}{#2}}}

\def\fpage#1{\begingroup
\voffset=.3in
\thispagestyle{empty}\begin{table}[b]\centerline{\footnotesize #1}
        \end{table}\endgroup}

\def\runninghead#1#2{\pagestyle{myheadings}
\markboth{{\protect\footnotesize\it{\quad #1}}\hfill}
{\hfill{\protect\footnotesize\it{#2\quad}}}}
\font\tenrm=cmr10
\font\tenit=cmti10
\font\tenbf=cmbx10
\font\bfit=cmbxti10 at 10pt
\font\ninerm=cmr9
\font\nineit=cmti9
\font\ninebf=cmbx9
\font\eightrm=cmr8

\def\qed{\hbox{${\vcenter{\vbox{                        
   \hrule height 0.4pt\hbox{\vrule width 0.4pt height 6pt
   \kern5pt\vrule width 0.4pt}\hrule height 0.4pt}}}$}}

\def\bsc{{\sc a\kern-6.4pt\sc a\kern-6.4pt\sc a}}       
\def\bflatex{\bf L\kern-.30em\raise.3ex\hbox{\bsc}\kern-.14em
T\kern-.1667em\lower.7ex\hbox{E}\kern-.125em X}

\begin{document}
\runninghead
{Knots and Numbers in $\phi^4$ Theory \ldots}
{\ldots\ to 7 Loops and Beyond}
\normalsize\textlineskip
\thispagestyle{empty}
\setcounter{page}{1}
\copyrightheading{}                     
\vspace*{0.88truein}
\fpage{1}
\centerline{\large\bf
Knots and Numbers in $\phi^4$ Theory to 7 Loops and Beyond\footnote{
Presented by D.\ J.\ Broadhurst,
at the AI--HENP 95 workshop, Pisa, April 1995}
\footnote{OU---Mainz collaboration, supported in part by HUCAM
(contract CHRX--CT94--0579)}}
\vspace*{0.37truein}
\centerline{\footnotesize
D.~J.~BROADHURST\footnote{
Email: D.Broadhurst@open.ac.uk}}
\vspace*{0.015truein}
\centerline{\footnotesize\it
Physics Department, Open University}
\baselineskip=10pt
\centerline{\footnotesize\it
Milton Keynes, MK7 6AA, UK}
\vspace*{10pt}
\centerline{\normalsize and}
\vspace*{10pt}
\centerline{\footnotesize
D.\ KREIMER\footnote{
On leave of absence from Department of Physics, University of Mainz, FRG}}
\vspace*{0.015truein}
\centerline{\footnotesize\it
Department of Physics, University of Tasmania}
\baselineskip=10pt
\centerline{\footnotesize\it
GPO Box 252C, Hobart, TAS 7001, Australia}
\vspace*{0.225truein}
\vspace*{0.21truein}
\abstracts{
We evaluate all the primitive divergences contributing to the 7--loop
$\beta$\/--function of $\phi^4$ theory, i.e.\ all 59 diagrams
that are free of subdivergences and hence give scheme--independent
contributions. Guided by the association of diagrams with knots, we obtain
analytical results for 56 diagrams. The remaining three diagrams, associated
with the knots $10_{124}$, $10_{139}$, and $10_{152}$, are evaluated
numerically, to 10 sf.
Only one satellite knot with 11 crossings is encountered and the
transcendental number associated with it is found.
Thus we achieve an analytical result for the 6--loop contributions,
and a numerical result at 7 loops that is accurate to one part in $10^{11}$.
The series of `zig--zag' counterterms, $\{6\zeta_3,\,20\zeta_5,\,
\frac{441}{8}\zeta_7,\,168\zeta_9,\,\ldots\}$, previously known
for $n=3,4,5,6$ loops, is evaluated to 10 loops, corresponding to
17 crossings, revealing that the $n$\/--loop zig--zag term is $4C_{n-1}
\sum_{p>0}\frac{(-1)^{p n - n}}{p^{2n-3}}$,
where $C_n=\frac{1}{n+1}{2n \choose n}$
are the Catalan numbers, familiar in knot theory.
The investigations reported here entailed intensive use of REDUCE,
to generate ${\rm O}(10^4)$ lines of code for multiple precision FORTRAN
computations, enabled by Bailey's MPFUN routines, running
for ${\rm O}(10^3)$ CPUhours on DecAlpha machines.}{}{}

\vspace*{1pt}\textlineskip      
\section{Introduction}          
\vspace*{-0.5pt}\noindent
At the AI--92 and AI--93 workshops,
I reported\cite{a92,a93} on a wide variety of multi--loop results.
Here, at AI--95,
I shall confine myself to a single topic, namely the very recent
results obtained by Dirk Kreimer and myself, at 7 loops, and beyond,
by combining knot theory with state--of--the--art symbolic and
numerical calculation. I refer you to the talks of my other collaborators
-- Pavel Baikov, Jochem Fleischer, John Gracey, Andrey Grozin,
Oleg Tarasov -- for reports of progress in other areas, and to the talk by
Dirk Kreimer, for discussion of the knot theory used here.

The investigation concerns the primitive divergences
contributing to the $\beta$\/--function
of $\phi^4$ theory, i.e.\ the counterterms that arise from diagrams
that are free of subdivergences and hence make scheme--independent
contributions to the $\beta$\/--function. A famous series of such
diagrams is the `zig--zag' series, beginning with\\
\setlength{\unitlength}{0.014cm}
\newbox\shell
\newcommand{\lbl}[3]{\put(#1,#2){\makebox(0,0)[b]{$#3$}}}
\newcommand{\dia}[1]{\setbox\shell=\hbox{\begin{picture}(180,200)(-90,-110)#1
\end{picture}}\dimen0=\ht
\shell\multiply\dimen0by7\divide\dimen0by16\raise-\dimen0\box\shell
}
\newcommand{\vtx}{\circle*{10}}
\dia{
\put(0,0){\circle{100}}
\put(-30,60){\line( 1,-2){60}}
\put( -5, 0){\line(-1, 0){65}}
\put(  5, 0){\line( 1, 0){65}}
\lbl{0}{-90}{Z(3)=6\zeta_3}
}
\dia{
\put(0,0){\oval(120,100)[r]}
\put(0,0){\oval(120,100)[l]}
\put(0,-50){\line(-1,2){55}}
\put(0,-50){\line( 1,2){55}}
\put(-80,0){\line(1,0){50}}
\put(-20,0){\line(1,0){40}}
\put( 30,0){\line(1,0){50}}
\lbl{0}{-90}{Z(4)=20\zeta_5}
}
\hspace{3mm}
\dia{
\put(0,0){\oval(150,100)[r]}
\put(0,0){\oval(150,100)[l]}
\put(-25,-50){\line( 1, 2){50}}
\put(-25,-50){\line(-1, 2){55}}
\put( 25, 50){\line( 1,-2){55}}
\put(-55,0){\line(-1,0){40}}
\put( -5,0){\line(-1,0){40}}
\put(  5,0){\line( 1,0){40}}
\put( 55,0){\line( 1,0){40}}
\lbl{0}{-90}{Z(5)=\frac{441}{8}\zeta_7}
}
\hspace{1cm}
\dia{
\put(0,0){\oval(200,100)[r]}
\put(0,0){\oval(200,100)[l]}
\put(  0, 50){\line(-1,-2){50}}
\put(  0, 50){\line( 1,-2){50}}
\put(-50,-50){\line(-1, 2){55}}
\put( 50,-50){\line( 1, 2){55}}
\put(-20,0){\line( 1,0){40}}
\put(-30,0){\line(-1,0){40}}
\put( 30,0){\line( 1,0){40}}
\put(-80,0){\line(-1,0){40}}
\put( 80,0){\line( 1,0){40}}
\lbl{0}{-90}{Z(6)=168\zeta_9}
}\\
To find the corresponding counterterms, one nullifies the 4 external
momenta, and cuts the horizontal line, to obtain planar two--point
diagrams, whose values (with unit external momentum, and loop--measure
${\rm d}^4k/\pi^2$) yield the indicated counterterms. The first two terms
are trivial to obtain: $Z(3)=W(3)$, $Z(4)=W(4)$, where
$W(n)=n C_{n-1}\zeta_{2n-3}$
is the value of any diagram obtained by cutting the $n$\/--loop
vacuum diagram consisting of a wheel with $n$ spokes\cite{wheel,ladder},
and $C_n\equiv\frac{1}{n+1}{2n \choose n}$ are the Catalan numbers.
Thereafter, the rational numbers in the zig--zag series differ from
those in the ladder\cite{ladder,BaU,UaD} series.
The 5--loop result was obtained in\cite{441} and verified in\cite{OUT};
the 6--loop result was obtained in\cite{OUT} and verified in\cite{168}.
The zig--zag diagrams, at 3, 4, and 5 loops, account for 44\%, 46\%, and
47\% of the terms in the MS $\beta$\/--function,
first calculated in\cite{oldbeta}, and corrected in\cite{newbeta}.
Moreover, they account for 100\%, 100\%, and 79\% of the scheme--independent,
primitive contributions. Hence there has long been an interest
in obtaining the $n$\/--loop result, which we shall infer from
numerical results, to $n=10$ loops.
We also compute {\it all\/} the primitive
terms in the $\beta$\/--function, to $n=7$ loops, and compare
them with expectations based on knot theory\cite{Dirk}.

\section{Expectations from Knot Theory}
\vspace*{-0.5pt}\noindent
The recent association\cite{Dirk} of knots with transcendental numbers, in
the counterterms of field theory, does not (yet, at least) allow
us to obtain the value of a diagram directly from the associated knot
(or knots).
It does, however, provide a powerful guide to the transcendentals that are
expected to occur, with rational coefficients, in the counterterm
obtained from any diagram that is free of subdivergences. (The
restriction to such primitive diagrams means that we do not need
to specify a scheme.) As explained in Dirk Kreimer's talk,
one can turn any $\phi^3$ diagram into a link diagram, to which a skein
relation is applied, yielding knots that are closures of positive braids.
There are very few such knots. Ignoring factor knots, only 10 knots
can be generated by diagrams with up to 7 loops. Of these, the 8 knots
with up to 10 crossings are
to be found in standard tables\cite{Jones,Rolfsen}, where they are known
as $3_1$, $5_1$, $7_1$, $8_{19}$, $9_1$, $10_{124}$, $10_{139}$, $10_{152}$,
with the first number indicating the number of crossings.
We denote the two positive--braid 11--crossing knots as $11_1$ and $11_{353}$,
for reasons that will become apparent when we identify the numbers that
they entail.

With the exceptions of the satellite knots
$10_{139}$, $10_{152}$, and $11_{353}$, all
positive knots with up to 11 crossings
are torus knots. The torus knot $(p,q)=(q,p)$, with $p$ and $q$ having no
common factor, is formed by a closed loop on a torus, winding round one
axis $p$ times, while it winds round the other $q$ times,
before rejoining itself. The $n$\/--loop ladder
diagram\cite{ladder,BaU,UaD}
yields the torus knot $(2n-3,2)$, with $2n-3$ crossings,
thereby giving us the entries
$N_1\sim\zeta_N$ in the knot--to--number dictionary\cite{Dirk}, with $N=2n-3$
crossings at the $n$\/--loop level.
Considering $\phi^4$ theory to be obtained from a $\phi^2\sigma$ coupling,
with a non--propagating $\sigma\sim\phi^2$, one concludes\cite{Dirk} that the
$n$\/--loop zig--zag counterterm will be a rational multiple of
$\zeta_{2n-3}$.

There are two more torus knots encountered at up to 7 loops:
$8_{19}=(4,3)$ and $10_{124}=(5,3)$. (With $\{(N,2)|N=3,5,7,9,11\}$,
they exhaust the mutually prime pairs $(p,q)$ such that the number of
crossing, $p q-\max(p,q)$, does not exceed 11.)
We shall identify the two double--sums associated with
these two non--zeta knots.

Exact expressions for the numbers associated with $10_{139}$ and $10_{152}$
are not yet available; we shall be content with numerical evaluation
of three diagrams whose skeinings yield combinations
of the three positive 10--crossing knots, $10_{124}$, $10_{139}$
and $10_{152}$.
We show that the triple--sum $F_{353}\equiv\sum_{l>m>n>0}l^{-3}m^{-5}n^{-3}$
is associated with $11_{353}$, and obtain analytical results for
all the diagrams that produce it.

Our methods involve high--precision numerical evaluations, yielding
numbers for which we seek analytical fits in knot--theoretically
prescribed search spaces. These search spaces correspond
to rational combinations of the transcendentals associated
with knots obtained by skeining\cite{Dirk} the link diagrams
that encode the topology of the intertwining of loop momenta in the
counterterm diagrams.

\section{Non--zeta Numbers and their Knots}
\vspace*{-0.5pt}\noindent
With such a plethora of diagrams to consider, it is convenient to
represent primitive counterterms by angular diagrams\cite{OUT,GPXT}, such as\\
\dia{
\put(0,0){\circle{100}}
\put(0,-50){\line(0,1){100}}
\put(0,50){\vtx}
\lbl{-60}{-5}{1}
\lbl{-10}{-5}{2}
\lbl{60}{-5}{3}
\lbl{0}{-90}{G(n_1,n_2,n_3)}
}\hfill
\dia{
\put(0,0){\oval(125,100)[r]}
\put(0,0){\oval(125,100)[l]}
\put(0,0){\oval(75,98)[r]}
\put(0,0){\oval(75,98)[l]}
\lbl{-72}{-5}{1}
\lbl{-25}{-5}{2}
\lbl{25}{-5}{3}
\lbl{72}{-5}{4}
\lbl{0}{-90}{M(n_1,\ldots,n_4)}
}\hfill
\dia{
\put(0,0){\circle{100}}
\put(0,50){\line(1,-2){40}}
\put(0,50){\line(-1,-2){40}}
\put(0,50){\vtx}
\lbl{-60}{-5}{1}
\lbl{-15}{-15}{2}
\lbl{0}{-45}{3}
\lbl{15}{-15}{4}
\lbl{60}{-5}{5}
\lbl{0}{-90}{C(n_1,\ldots,n_5)}
}\hfill
\dia{
\put(0,0){\circle{100}}
\put(0,50){\line(1,-2){40}}
\put(0,50){\line(-1,-2){40}}
\put(-40,-30){\vtx}
\lbl{-60}{-5}{1}
\lbl{-15}{-15}{2}
\lbl{0}{-45}{3}
\lbl{15}{-15}{4}
\lbl{60}{-5}{5}
\lbl{0}{-90}{D(n_1,\ldots,n_5)}
}\\
where, for example, the arguments of $G(n_1,n_2,n_3)$ specify the number
of dots to be placed on the three lines, and all dots are connected
to an origin (which is not shown). Then one uses the Gegenbauer
expansion for the massless propagator
with $L$ dots, which is know in closed form\cite{ladder}, for
arbitrary $L$. The radial
integrations can be done analytically. (We used REDUCE\cite{red}.)
The angular
integrations yield, in the case of the first topology, the
triangle function $\Delta(l,m,n)$, which is 1, or 0,
according as $(l+m+n+1)/2$ is, or is not, an integer greater than
all of the Gegenbauer expansion indices, $\{l,m,n\}$, introduced
in expanding the three `dotted' propagators. For this topology,
we obtain a finite number of infinite triple--sums:
\begin{eqnarray*}
\!G(\alpha,\beta,\gamma)\!&=&\!
\sum_{i,j,k}
{2\alpha-i\choose\alpha}
{2\beta -j\choose \beta}
{2\gamma-k\choose\gamma}
\frac{(i+j+k)!}{i!j!k!}
\!\!\sum_{l,m,n}\!\!
\frac{\Delta(l,m,n)}{l^{2\alpha-i}m^{2\beta-j}n^{2\gamma-k}}
\\
&\times&
\left[
\left(\frac{2}{l+m+n-1}\right)^{i+j+k+1}+
\left(\frac{2}{l+m+n+1}\right)^{i+j+k+1}\right]\,.
\end{eqnarray*}
Manipulating the sums\cite{OUT}
$F_{ab}\equiv\sum_{l>m>0}l^{-a}m^{-b}$,
$F_{abc}\equiv\sum_{l>m>n>0}l^{-a}m^{-b}n^{-c}$,
we were able to obtain
analytical results for all diagrams of type G, up to 7 loops,
in terms of zetas and a {\it single\/} non--zeta level--11 transcendental
\newcommand{\Df}[2]{\mbox{$\frac{#1}{#2}$}}
\[K_{353}\equiv\frac{G(3,2,0)-583\zeta_{11}}{40}
=\Df{3}{5}(F_{353}-\zeta_3F_{53})
+\Df{69}{40}\zeta_{11}-\zeta_5\zeta_3^2
=0.2019352393\]
associated with a unique, 11--crossing, positive--4--braid, satellite knot,
which we consequently denote as $11_{353}$.
In knot theory, it is identified by the braid word
$\sigma_1^2\sigma_2^2\sigma_1^{}\sigma_3^{}\sigma_2^3\sigma_3^2$.
A REDUCE program, running for 2 days on a DecAlpha, established
its uniqueness, by computation of $3^{11}=177\,147$
Jones polynomials\cite{Jones}.

To find the new transcendentals entailed by the non--zeta torus
knots $8_{19}=(4,3)$ and $10_{124}=(5,3)$, we studied the 4--fold sums
generated by angular diagrams of type M, evaluating all such diagrams,
up to 7 loops. Thanks to powerful methods of accelerated convergence,
multiple--precision\cite{mpf} FORTRAN, and efficient integer--relation
search routines\cite{mpf}, it was easy to find, at 15 sf, and to verify,
up to 45 sf, the analytical form of the sole level--8 non--zeta
transcendental
\[K_{53}\equiv7\zeta_5\zeta_3-\Df{1}{36}M(2,1,1,0)
=\Df{1}{5}(29\zeta_8-12F_{53})
=5.733150251\]
whose appearance we associate with the knot $8_{19}=(4,3)$.
(The probability of an accidental fit is of order $10^{-30}$.)
This is the only non--zeta number appearing in the expansion, to level
9, of the master two--loop diagram\cite{1440,Z2S6}, where it enters
via\cite{1440}
\[U_{6,2}\equiv\sum_{n>m>0}\frac{(-1)^{n-m}}{n^6m^2}
=\Df{747}{256}\zeta_8-\Df{3}{2}\zeta_5\zeta_3-\Df{3}{16}K_{53}
=-0.01476857328\,.\]

At level 10, we obtained the knot $10_{124}=(5,3)$ from skeining $M(2,1,1,1)$,
and found, at high precision, a fit to this diagram
in the search space $\{F_{73},\zeta_{10},\zeta_5^2,\zeta_7\zeta_3\}$.
So as to simplify the appearance of other results that involve
$10_{124}$, we find it convenient to express the result as follows:
\[K_{73}\equiv\frac{90\zeta_5^2-21\zeta_7\zeta_3-\frac{1}{36}M(2,1,1,1)}{7}
=\frac{94F_{73}+982\zeta_5^2-793\zeta_{10}}{28}
=9.388141968\,.\]

We thus identify
$\{\zeta_3,\zeta_5,\zeta_7,K_{53},\zeta_9,K_{73},\zeta_{11},K_{353}\}$
(and their products, corresponding to factor knots)
as the only transcendentals that may appear in primitive diagrams whose
skeinings do not involve the satellite knots $10_{139}$ and $10_{152}$.
By methods too laborious to recount here, we succeeded in evaluating
all primitive diagrams of types M, C and D, up to 7 loops, to 50 sf,
and (to our great delight) found integer relations,
of the expected form, i.e.\ with no trace of $10_{139}$
or $10_{152}$. (The probability of accidental fitting is typically
$10^{-20}$, since the search routines need less than 30 sf, to produce
results which hold to 50 sf.)

Our tables of new analytical results are too long to
reproduce here. They have a feature that is notable: every planar diagram,
and every diagram with 4--point (or lower) vertices
(whether planar or non--planar),
is a rational combination of transcendentals at the {\it same\/} level.
Only for diagrams that are non--planar and also contain
5--point (or higher) vertices do we (sometimes) observe level--mixing.
For example, the planar diagrams $G(3,2,0)$,
$G(4,1,0)$, $C(3,0,0,0,1)$, $C(1,0,2,0,1)$, $D(3,0,0,0,1)$,
$D(1,0,2,0,1)$ are pure level--11, whilst $G(2,2,1)$ and $G(3,1,1)$
are non--planar diagrams, with 6--point vertices, and mix
levels 6 through 10.

\section{Primitive $\phi^4$ Counterterms, to 7 Loops and Beyond}
\vspace*{-0.5pt}\noindent
Encouraged by the foregoing, we sought to evaluate {\it all\/} the primitive
$\phi^4$ counterterms, up to 7 loops, and the zig--zag counterterms
to 10 loops. This entails more complex angular diagrams,
from which REDUCE\cite{red} generates summands,
thousands of lines long, processed by
MPFUN\cite{mpf} routines that evaluate sums over as many as
8 integers, with angular factors that may involve squares of 6--j
symbols\cite{OUT,Nickel}. Thanks to about $10^3$ CPUhours on
a pair of DecAlphas, and a commensurate human effort, to develop
methods of accelerated convergence, we succeeded in finding
high--precision, knot--theory inspired, analytical fits to all but
3 of the 59 diagrams with up to 7 loops
(generated automatically, by a purpose--built
Wick--contracting program). At 3 loops, there is one diagram, $Z(3)=W(3)$.
At 4 loops, there is one diagram, $Z(4)=W(4)$. At 5 loops, there are
3 diagrams, reduced by conformal transformation to 2 distinct
values: $Z(5)=\frac{63}{80}W(5)$ and $M(1,1,1,0)=[W(3)]^2$.
At 6 loops, there are 10 diagrams, reduced by conformal transformations to
5 distinct values:
$Z(6)=\frac23W(6)$, $W(3)W(4)$,
$M(2,1,1,0)$, $M(1,1,1,1)=12W(3)W(4)-16M(2,1,1,0)$, and
$D(2,0,0,0,1)=\frac{1063}{9}\zeta_9+8\zeta_3^3$. For the coefficients,
$\beta^{\rm NS}_n$, of the primitive (i.e.\ no--subdivergence)
terms in $\beta(g)\equiv\mu^2{\rm d}g/{\rm d}\mu^2=\sum_n\beta_n (-g)^{n+1}$,
with an interaction ${\cal H}_{\rm int}=(4\pi)^2g\,\phi^4/4!$, we obtain
$\beta_3^{\rm NS}=6\zeta_3$,
$\beta_4^{\rm NS}=60\zeta_5$,
$\beta_5^{\rm NS}=\frac{1323}{2}\zeta_7+126\zeta_3^2$, and the new
6--loop result
\[\beta_6^{\rm NS}=\Df{23056}{3}\zeta_9
+384\zeta_3^3+4512\zeta_5\zeta_3-336K_{53}=
12\,065.365\,126\,645\,.\]

At 7 loops, there are 44
diagrams, reduced to 12 distinct values, by intensive use of
conformal transformations, and re--interpretation of $x$\/--space variables
as $p$\/--space variables, which transforms a planar diagram into
one whose lines cross those of the original\cite{1440}.
The weights and values of the 12 numbers in $\beta_7^{\rm NS}=
\sum_n w_n B_n$ are
\[\begin{array}{rll}
n &w_n&B_n\\
1 & \frac{27}{4} & 216 \zeta_3^3 \\
2 & 18   & 400 \zeta_5^2 \\
3 & 12   & \frac{33759}{64}\zeta_{11} \\[1pt]
4 & 108  & \frac{4895}{64}\zeta_{11}+120\zeta_5\zeta_3^2+85K_{353} \\
5 & 81   & 420 \zeta_7\zeta_3 - 200 \zeta_5^2 \\
6 & \frac{27}{2} & 183.032\,420\,030\,498\,901\,717\,011\,912\,3(1) \\
7 & 108  & \frac{67925}{192}\zeta_{11}+20\zeta_5\zeta_3^2-15K_{353} \\[1pt]
8 & 108  & \frac{18601}{48}\zeta_{11}-28\zeta_5\zeta_3^2-48K_{353} \\
9 & 18   & 216.919\,375\,55(6) \\
10& 30   & 450 \zeta_5^2 - 189 \zeta_7\zeta_3 \\
11& 54   & \frac{1323}{4}\zeta_7\zeta_3 \\
12& 12   & 200.357\,566\,43(2)
\end{array}\]
where $B_{1,2,11}$ are factor knots;
$B_3=Z(7)$ is the 7--loop zig--zag;
$B_{4,7,8}$ entail $11_{353}$;
\newpage\noindent
$B_{5,10}$ are combinations of 10--crossing factor knots;
$B_{6,9,12}$ entail $10_{124}$, $10_{139}$, $10_{152}$.

We regard the fit between knots and numbers, in
59 $\phi^4$ counterterms, to 7 loops, as strongly indicative
of the kinship of knot theory and field theory.
We lack only a single, 6--element, integer--relation,
between $\{B_6,B_9,B_{12}\}$ and $\{K_{73},\zeta_7\zeta_3,\zeta_5^2\}$,
which the accuracy achieved for the fearsome 8--fold sums in $B_{9,12}$ is
insufficient to reveal. We have found such relations
for diagrams with 5-- and 6--point vertices.

{}From $Z(7)=\frac{1023}{1792}W(7)$ and
$Z(8)=\frac12W(8)$,
we infer the complete zig--zag series
\[Z(n)=4C_{n-1}\sum_{p=1}^\infty\frac{(-1)^{p n - n}}{p^{2n-3}}
=\left\{\begin{array}{rl}
          4C_{n-1}\zeta_{2n-3}\,,&\mbox{for even $n$\,,}\\
(4-4^{3-n})C_{n-1}\zeta_{2n-3}\,,&\mbox{for odd $n$\,,}
\end{array}\right.\]
which we have verified, to high precision, up to $n=10$ loops.
(An all--order proof is lacking, as yet.)
This series gives a convergent contribution to the $\beta$\/--function,
and hence is of decreasing importance at higher orders, where we observe
the growth
\[
\beta_5^{\rm NS}/\beta_4^{\rm NS}=13.647456527;\quad
\beta_6^{\rm NS}/\beta_5^{\rm NS}=14.209833853;\quad
\beta_7^{\rm NS}/\beta_6^{\rm NS}=15.371460754\]
which we shall compare with asymptotic expectations\cite{Lipatov,KST} in
a more detailed paper.

I conclude by thanking David Bailey and Andrey Grozin for
their generous help, and by noting, ruefully, my failure to confute Dirk
Kreimer's exciting new ideas\cite{Dirk}.

\nonumsection{References}

\end{document}